\documentclass[12pt,aps,prb,preprint]{revtex4}
\usepackage{amsmath}
\usepackage{graphicx}
\begin{document}
\title{

Comment on ``Quantum mechanical versus semiclassical tunneling and decay times" }
\author{Robert L. Intemann}
\affiliation{Temple University, Department of Physics,
Philadelphia, PA 19122} \email{intemann@temple.edu}
\date{\today}

\maketitle
Shegelski, Kavka, and Hynbida\cite{shegelski} have shown how to calculate the lifetime of a particle initially localized in a potential well exactly quantum mechanically by employing a heuristic expression with a simple interpretation for the lifetime.  Their method allows for the inclusion of a bound state, and their results for tunneling through a centrifugal barrier demonstrate the major role that bound states play in determining the lifetime.

However, a drawback to their procedure, especially from the point of view of undergraduate student accessibility, is that it leads to numerical integrations over time that these authors acknowledge are very challenging due to rapid oscillations in the integrands and the need to introduce an artificial cutoff for their control.  These difficulties are inherent to their approach and are not ameliorated by considering simpler barriers such as a square barrier or a repulsive $\delta$-function barrier.

In this note we show that by choosing a slightly different, but equally appropriate, heuristic expression for the lifetime we can greatly simplify the analysis and arrive at analytical results requiring only a single final numerical integration that can be performed quickly and without difficulty.  We obtain results for the centrifugal barrier which are exact for the case where the system has no bound states.  For the case where a bound state is present we make one approximation which, as our results demonstrate, is relatively mild and preserves very well the essential feature of the exact calculation of Shegelski, et al., viz., the influence that the presence of a bound state has on the lifetime.

The expression that we choose to characterize the lifetime is $<t^2>^{1/2}$ as defined by

\begin{equation}
<t^2>\equiv
{\displaystyle \int_{0}^{\infty}t^2\,\Delta P_{in}(t)\, dt\over
\displaystyle \int_{0}^{\infty}\,\Delta P_{in}(t)\, dt},\label{one}
\end{equation}
\begin{equation}
P_{in}(t)=\int_{0}^{a}\,\mid\Psi(x,t)\mid^{\,2}\,dx,
\end{equation}
where, as in Ref. (1),
\begin{equation}
\Delta P_{in}(t)=\left[P_{in}(t)-P_{in}(\infty)\right]\theta(\left[P_{in}(t)-P_{in}(\infty)\right]),\label{three}
\end{equation}
and the step function is defined by $\theta(x)=0$ for $x<0$ and $\theta(x)=1$ for $x>0$.

The wave function at time $t$ is a linear superposition of bound and unbound eigenstates of the Hamiltonian,
\begin{eqnarray}
\Psi(x,t) & = & C_{b}\phi_{b}(x)\,e^{-iE_{b}t/\hbar}+\sum_{k}C_{k}\phi_{k}(x)\,e^{-iE_{k}t/\hbar},\\
          & \equiv & \Psi_{b}(x,t)+\Psi_{u}(x,t),
\end{eqnarray}
where, as in Ref. 1, we have assumed the presence of only one bound state.

Thus,
\begin{equation}
P_{in}(t)=\int_{0}^{a}\mid\Psi_{b}(x,t)\!\mid^{\,2}dx+\int_{0}^{a}\mid\Psi_{u}(x,t)\!\mid^{\,2}dx+2Re\int_{0}^{a}\Psi_{b}^{\ast}(x,t)\Psi_{u}(x,t)\,dx.\label{six}
\end{equation}
The first term on the right of Eq.\ (\ref{six}) is simply $P_{in}(\infty)$ so that
\begin{equation}
 P_{in}(t)-P_{in}(\infty)=\int_{0}^{a}\mid\Psi_{u}(x,t)\!\mid^{\,2}dx+2Re\int_{0}^{a}\Psi_{b}^{\ast}(x,t)\Psi_{u}(x,t)\,dx.\label{DeltaP}
\end{equation}
The contribution from the second term on the right of Eq.\ (\ref{DeltaP}), the interference term between the bound and unbound states,we now argue is relatively unimportant. To see this we need only consider the orthogonality property of the bound and unbound energy eigenfunctions,
\begin{equation}
\int_{0}^{L}\phi_{b}^{\ast}(x)\phi_{k}(x)\,dx=\int_{0}^{a}\phi_{b}^{\ast}(x)\phi_{k}(x)\,dx+\int_{a}^{L}\phi_{b}^{\ast}(x)\phi_{k}(x)\,dx=0.
\end{equation}
As illustrated in Fig. 1, the region to the right of $x=a$ is a classically forbidden region for the bound state, and its wave function decreases to zero exponentially there.  Thus the integral over the classically forbidden region ($a\leq x\leq L$) is expected to be quite small, and the integral over the allowed region ($0\leq x\leq a$) must, in turn, be quite small also.  On this basis we shall neglect the interference term and henceforth represent $\Delta P_{in}(t)$ by
\begin{equation}
\Delta P_{in}(t)=\int_{0}^{a}\mid\Psi_{u}(x,t)\!\mid^{\,2},
\end{equation}
having noted that in this approximation $P_{in}(t)-P_{in}(\infty)\ge 0$ so that the step function in Eq.\ (\ref{three}) is no longer needed.

We now proceed to the calculation of $<t^2>$.  From Eq. (10) of Ref. 1 we have\cite{units}
\begin{equation}
\Psi_{u}(x,t)=\int_{0}^{\infty}dk\,\phi(k)\sin(qx)\,e^{-i(ka)^2(t/t_{0})},
\end{equation}
with
\begin{equation}
\phi(k)=2\sqrt{2a}\,\frac{\sin(qa)}{[\pi^2-(qa)^2]}\frac{k^2}{f^2(k)},
\end{equation}

\vspace{0.5cm}
\noindent and $q\equiv q(k)=\sqrt{k^2-2V_{0}}$ ($V_{0}<0$ for a well) as $L\rightarrow\infty$ and $0\leq x\leq a$. The function $f^2(k)$ is defined by Eq. (18) of Ref. 1. For convenience, we record here its actual form for a centrifugal barrier with $\ell=1$:
\begin{equation}
f^2(k)=\frac{1}{\kappa^2 a^2}\left[(1+\kappa^2)\alpha^2\cos^2\alpha+(1-\kappa^2+\kappa^4)\sin^2\alpha+\alpha\sin(2\alpha)\right],
\end{equation}
with $\kappa=ka$ and $\alpha=qa$.
Then, beginning with the $x$-integration,
\begin{eqnarray}
\Delta P_{in}(t) & = & \int_{0}^{\infty}dk\int_{0}^{\infty}dk^{\,\prime}\,\phi(k)\phi(k^{\,\prime})\,
e^{i[(k^{\,\prime}a)^2-(ka)^2](t/t_{0})}\int_{0}^{a}dx\sin(qx)\sin(q^{\,\prime}x),\\
              &  =  & \int_{0}^{\infty}dk\int_{0}^{\infty}dk^{\,\prime}\,\Phi(k,k^{\,\prime})\,e^{if(k,k^{\,\prime})t},
\end{eqnarray}
where
\begin{equation}
\Phi(k,k^{\,\prime})=\phi(k)\phi(k^{\,\prime})\chi(q,q^{\,\prime}),
\end{equation}

\begin{equation}
\chi(q,q^{\,\prime})=\frac{q^{\,\prime}\cos(q^{\,\prime}a)\sin(qa)-q\cos(qa)\sin(q^{\,\prime}a)}{(q^2-q^{\,\prime2})},
\end{equation}
and
\begin{equation}
f(k,k^{\,\prime})=\left[(k^{\,\prime}a)^2-(ka)^2\right]/t_0.
\end{equation}

Consider first the numerator in Eq.\ (\ref{one}).  We observe that $\Phi(k,k^{\,\prime})$ is invariant under $k\leftrightarrow k^{\,\prime}$ and is an even function of both $k$ and $k^{\,\prime}$. These symmetries allow us to write
\begin{subequations}
\begin{eqnarray}
\int_{0}^{\infty}dt\, t^2\Delta P_{in}(t) & = & \frac{1}{2}\int_{-\infty}^{\infty}dt\, t^2\Delta P_{in}(t)\\
                                        & = & \frac{1}{4}\int_{0}^{\infty}dk\int_{-\infty}^{\infty}dk^{\,\prime}
                                        \Phi(k,k^{\,\prime})\int_{-\infty}^{\infty}dt\, t^2\,e^{if(k,k^{\,\prime})t}\\
                                        & = &\frac{1}{4}\int_{0}^{\infty}dk\int_{-\infty}^{\infty}dk^{\,\prime}
                                        \frac{\Phi(k,k^{\,\prime})}{k\,k^{\,\prime}}\frac{\partial^2}{\partial k\,\partial k^{\,\prime}}\int_{-\infty}^{\infty}dt\,e^{if(k,k^{\,\prime})t},\\
                                        & = & \frac{\pi}{2}\int_{0}^{\infty}dk\int_{-\infty}^{\infty}dk^{\,\prime}
                                        \frac{\Phi(k,k^{\,\prime})}{k\,k^{\,\prime}}\frac{\partial^2}{\partial k\,\partial k^{\,\prime}}\delta\left[f(k,k^{\,\prime})\right],\label{dirac}\\
                                        & = & \frac{t_0}{2a^2}\frac{\pi}{2}\int_{0}^{\infty}dk\int_{-\infty}^{\infty}dk^{\,\prime}
                                        \frac{\Phi(k,k^{\,\prime})}{k\,k^{\,\prime}}\frac{\partial^2}{\partial k\,\partial k^{\,\prime}}\frac{\left[\delta(k-k^{\,\prime})+
                                        \delta(k+k^{\,\prime})\right]}{\mid k\mid},
\end{eqnarray}
\end{subequations}
where the Dirac $\delta$-function was introduced in Eq.\ (\ref{dirac}) through its Fourier integral representation.
Integrating by parts with respect to $k$ and $k^{\,\prime}$, and noting that the integrated terms vanish,we obtain
\begin{equation}
\int_{0}^{\infty}dt\, t^2\Delta P_{in}(t)=\frac{t_0}{2a^2}\frac{\pi}{2}\int_{0}^{\infty}\frac{dk}{k}\int_{-\infty}^{\infty}dk^{\,\prime}\frac{\partial^2\Psi(k,k^{\,\prime})}{\partial k\,\partial k^{\,\prime}}\left[\delta(k-k^{\,\prime})+\delta(k+k^{\,\prime})\right],
\end{equation}
where $\Psi(k,k^{\,\prime})\equiv\Phi(k,k^{\,\prime})/k\,k^{\,\prime}$.

Finally, then
\begin{equation}
\int_{0}^{\infty}dt\, t^2\Delta P_{in}(t)=\pi\int_{0}^{\infty}\frac{dk}{k}\left.\frac{\partial^2\Psi(k,k^{\,\prime})}{\partial k\,\partial k^{\,\prime}}\right|_{k^{\,\prime}=k},\label{numerator}
\end{equation}
where we have made use of the definition $t_0=2a^2$ in arriving at the last line.
A similar calculation for the denominator in Eq.\ (\ref{one}), but without the need to do any integrations by parts, results in
\begin{equation}
\int_{0}^{\infty}dt\,\Delta P_{in}(t)=\pi\int_{0}^{\infty}\frac{dk}{k}\,\Phi(k,k).\label{denominator}
\end{equation}

To evaluate $<t^2>$, the remaining integration in Eqs.\ (\ref{numerator}) and\ (\ref{denominator}) must be performed numerically, but in each case only a single integration is required. It can be done quickly and without difficulty using {\em Mathematica} or {\em Maple}.

To compare our results with the exact results of Ref. 1, we recall that for a system that decays exponentially, $<t^2>^{1/2}=\sqrt{2} <t>$.  Although the behavior of $\Delta P_{in}(t)$ is more complicated than a simple exponential, its overall effect on the lifetime should be roughly ``exponential-like".  Thus, if we {\em define} the average lifetime $<\bar{t}>$ for our approach by
\begin{equation}
<\bar{t}>=\frac{<t^2>^{1/2}}{\sqrt{2}},
\end{equation}
we expect that $<\bar{t}>$ will differ very little from $<t>$ the average lifetime as defined in Ref. 1.  This will enable us to make a direct comparison with the results presented in Fig. 2 of Ref. 1 by making a similar plot using $<\bar{t}>$.  For plotting purposes we introduce a dimensionless lifetime $\bar{\tau}=<\bar{t}>/t_0$ and plot $\bar{\tau}$ as a function of $<e>$ as defined by Eq. (20) of Ref. 1.  Our results are shown in Fig. 2 along with the WKB lifetime computed from Eq. (28) of Ref. 1.

A careful comparison of these two plots leads one to conclude that there is no perceptible difference between the right-hand half of the dotted curves.  Since both calculations are exact for the regime in which no bound state is present, we conclude that for comparison purposes $<\bar{t}>$ and $<t>$ may be considered to be the same.  With regard to the left-hand half of the dotted curves, there is some reduction in the values of  the $<\bar\tau>$ of our results and a sharper bend upward as one approaches close to the point where the bound state disappears.  It is in this region that one expects the interference term, which we have neglected, to exert its greatest influence.  But qualitatively, our results for the bound state regime are remarkably similar to the exact results of Ref. 1.  Thus, they too represent a significant improvement on the WKB approximation while, at the same time, maintaining a much greater simplicity of computation.

Finally, for those who wish to avoid the use of the Dirac $\delta$-function, we offer an alternate method for calculating the numerator and denominator in Eq.\ (\ref{one}).  It consists of inserting a convergence factor $e^{-\alpha t}$ into the original expressions for these quantities so that the $t$-integration can be carried out immediately (along with the $x$-integration).  The $k^{\,\prime}$-integration is then performed using the method of residues after which one can safely set $\alpha\rightarrow 0$.  There then remains only the $k$-integration that, once again, must be done numerically.

{\em Problem 1:}
Apply the convergence factor/residue method to the calculation of the denominator in Eq.\ (\ref{one}) and show that the result obtained is identical to that given by Eq.\ (\ref{denominator}).

{\em Problem 2:}
Apply the convergence factor/residue method to the calculation of the numerator in Eq.\ (\ref{one}) and show that the result obtained is equivalent to Eq.\ (\ref{numerator}) by demonstrating that it yields exactly the same numerical result as Eq.\ (\ref{numerator}).

\newpage
\begin{figure}[h]
\begin{center}
\includegraphics{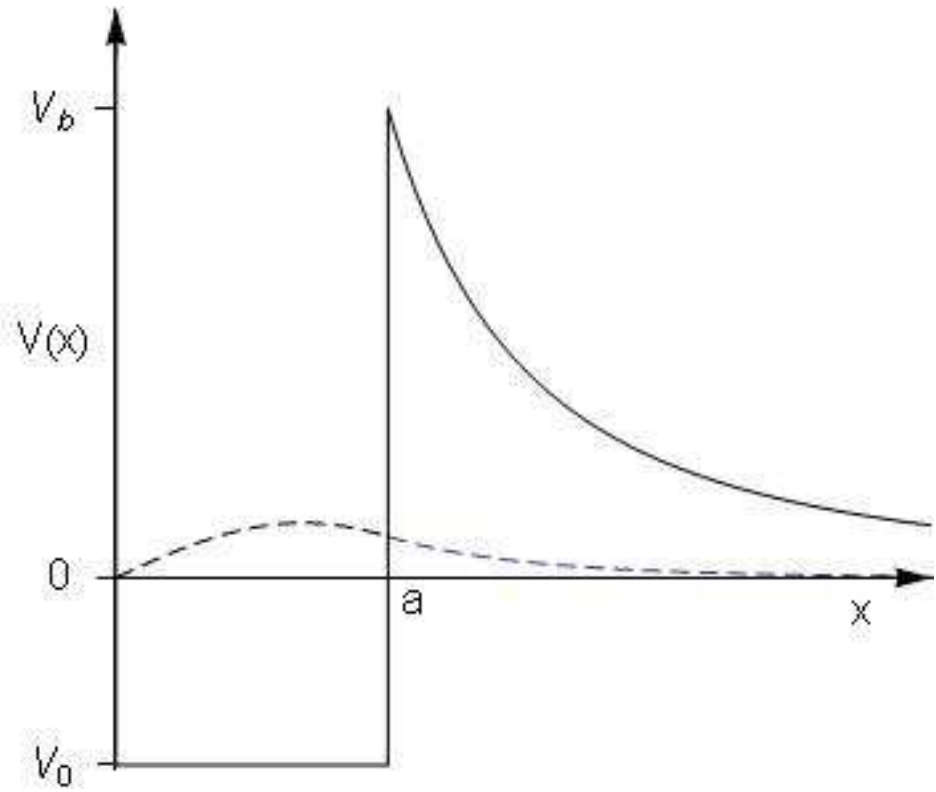}
\caption{\label{fig:potential}The potential for the centrifugal barrier problem.  The dashed curve illustrates a typical bound state wave function for the case where a bound state is present.}
\end{center}
\end{figure}
\newpage
\begin{figure}[h]
\begin{center}
\includegraphics{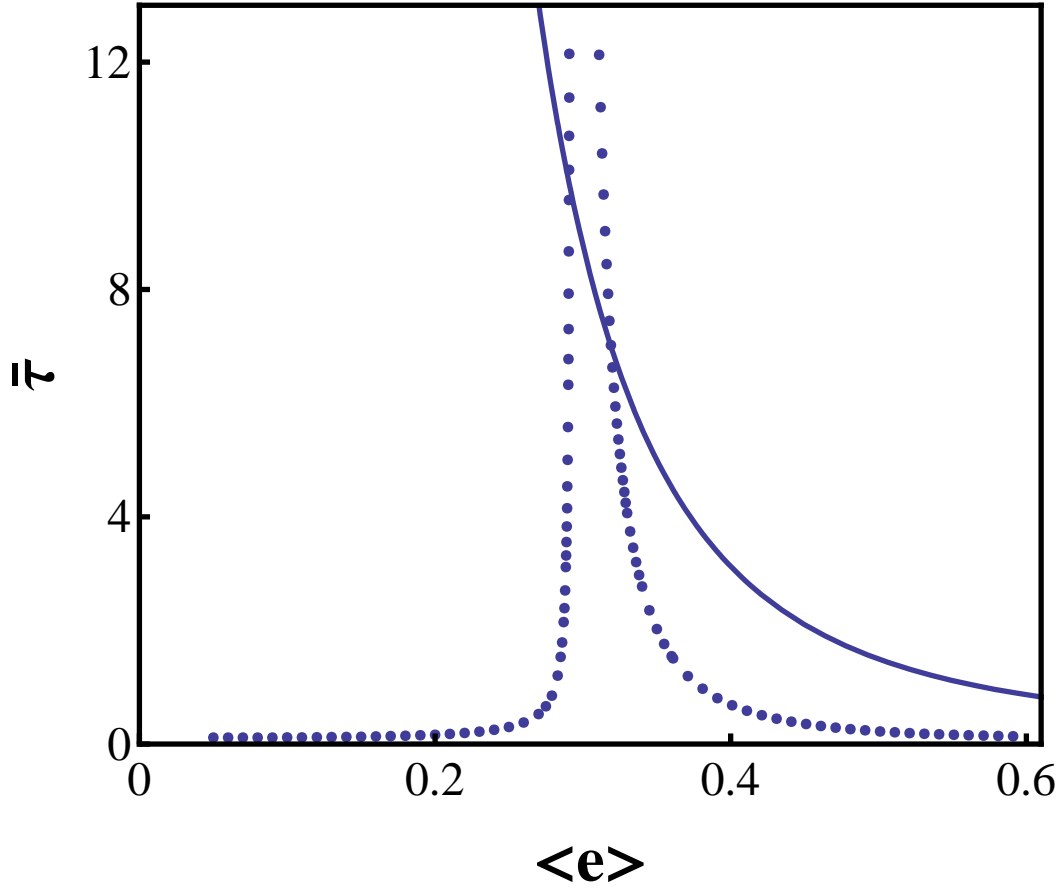}
\caption{\label{fig:lifetime}The tunneling time as a function of the dimensionless energy $<e>=<E>/V_b$, where $<E>$ is the expectation value of the particle's energy and $V_b$ is the height of the centrifugal barrier.  The solid circles are the results of the present work and the continuous curve represents the WKB approximation.}
\end{center}
\end{figure}

\begin{thebibliography}{2}
\bibitem{units}{We have set $m=\hbar=1$ at this point for convenience.}
\end{thebibliography}
\end{document}